\documentclass[%
 reprint,
 superscriptaddress,
 amsmath,amssymb,
 aps,
 prl,
nolongbibliography,
]{revtex4-2}

\usepackage{graphicx}
\usepackage{dcolumn}
\usepackage{bm}
\usepackage{hyperref}
\usepackage[dvipsnames]{xcolor}
\hypersetup{
    colorlinks = true,
    linkcolor = blue,
    anchorcolor = blue,
    citecolor = blue,
    filecolor = blue,
    urlcolor = black
}

\newcommand{\added}[1]{#1}

\begin{document}

\preprint{APS/123-QED}

\title{Parity Breaking at Faceted Crystal Growth Fronts during Ice Templating}

\author{Kaihua Ji}
 \email[]{ji5@llnl.gov}
\affiliation{%
Department of Physics and Center for Interdisciplinary Research on Complex Systems, Northeastern University, Boston, Massachusetts 02115, USA
}%
\affiliation{%
Lawrence Livermore National Laboratory, Livermore, 94550, CA, USA
}%
\author{Alain Karma}%
 \email[]{a.karma@northeastern.edu; Corresponding author}
\affiliation{%
Department of Physics and Center for Interdisciplinary Research on Complex Systems, Northeastern University, Boston, Massachusetts 02115, USA
}%

\date{\today}

\begin{abstract}
Directional solidification of water-based solutions has emerged as a versatile technique to template hierarchical porous materials, but this nonequilibrium process remains incompletely understood. Here we use phase-field simulations to shed light on the mechanism that selects the growth direction of the lamellar ice structure that templates those materials. Our results show that this selection can be understood within the general framework of spontaneous parity breaking, yielding quantitative predictions for the tilt angle of lamellae with respect to the thermal axis. The results provide a theoretical basis to interpret a wide range of experimental observations. 
\end{abstract}

\maketitle


Directional solidification of water-based solutions or slurries often produces intricate faceted microstructural patterns. While the non-faceted patterns commonly observed in alloy solidification (e.g., cellular and dendritic structures) are relatively well understood, much less is known about faceted crystal growth. This faceted growth is crucial to ice templating, also known as freeze casting—a processing technique that uses ice crystals as templates to produce porous materials with hierarchical architectures \cite{Deville2006FreezingComposites,Wegst2010BiomaterialsCasting,Donius2014SuperiorCasting,scotti2018freeze,wegst2024freeze}. This technique has been utilized to manufacture a variety of materials, such as polymers, ceramics, and metals, demonstrating potential for applications in biomedicine \cite{Riblett2012Ice-TemplatedGrowth,Bozkurt2012TheNerves,Deville2006FreezeEngineering,Mohan2018FluorescentModel,yin2019freeze}, energy generation and storage \cite{Han2019AnisotropicAbsorption,Qiu2022ExcellentFoams,Shao2020FreezeApplications}. Notably, ice crystals frequently tilt relative to the externally imposed temperature gradient during growth, and the resulting scaffolds often exhibit lamellar cell walls with unilateral surface features oriented toward the hot side of the temperature gradient~\cite{Deville2013Time-lapseSuspension,kamm2023x}. 

In a recent study \cite{Yin2023HierarchicalTemplating}, phase-field (PF) modeling was used to investigate the formation of those unilateral features ranging from secondary ridges forming perpendicularly to the primary ice lamellae to more exotic features resembling living forms. While this study explained a number of experimental observations in water-based solutions, it left the open question of how the tilt angle $\gamma$ of primary ice lamellae with respect to the temperature gradient axis is dynamically selected, which is the main focus of the present Letter. We show that tilt selection can be understood within the general framework of parity breaking at crystal growth fronts, which has been used previously to interpret observations of drifting periodic patterns in liquid-crystal systems \cite{Simon1988SolitarySolidification,Bechhoefer1989DestabilizationInterface} and  
lamellar eutectic growth \cite{Karma1987BeyondGrowth,Faivre1989SolitaryEutectics,Kassner1990ParityGrowth,Kassner1991SpontaneousStructures}. However, unlike in those systems where interface growth is nearly isotropic and close to local equilibrium, ice growth is strongly anisotropic and far from equilibrium due to the presence of facets. As a result, a full understanding of tilted pattern selection requires to consider both ``spontaneously broken'' parity, leading to asymmetrical growth even when the fast $a$-axis of ice crystals ($\left[11\bar{2}0\right]$ growth direction) is aligned parallel to the temperature gradient \cite{Yin2023HierarchicalTemplating,chen2025asymmetric}, and ``externally broken'' parity when the $a$-axis is at a finite misorientation $\gamma_0\ne 0$ with respect to the temperature gradient axis, which is generally the case in experiments where ice crystal grains extending several lamellar spacings have a small misorientation of a few degrees \cite{Deville2013Time-lapseSuspension,kamm2023x}. As theoretically expected, externally broken parity for finite $\gamma_0$ leads to the existence of two steady-state branches of drifting lamellar structures with different tilt angles. Our results characterize the relationships between tilt angle and misorientation for those two branches and further show that the pattern with the smallest tilt angle, which has facets tilted towards the cold side of the temperature gradient, is dynamically selected through a competitive growth process. This selection provides a natural explanation for the common observation that surface features of templated structures are oriented toward the opposite hot side. 

We use in what follows a sharp-interface formulation to list the equations and physical parameters controlling the directional solidification of dilute water-based binary solutions.
The PF model used in \cite{Yin2023HierarchicalTemplating} and the present study to numerically solve this free-boundary problem are given in a companion article \cite{Ji2025QuantitativeTemplating}, which provides a detailed exposition of the convergence of the model as a function of the diffuse-interface thickness and additional results. This convergence is made non-trivial by the presence of connected faceted and non-faceted regions of the solid-liquid interface that have dramatically different growth kinetics. Fast growth of atomically rough regions in the basal plane is weakly anisotropic and approximately in local equilibrium while faceted growth along the $c$-axis perpendicular to this plane is sluggish and far from equilibrium (i.e., with a large kinetic undercooling on the facets). The PF formulation quantitatively models ice crystal growth by smoothly interpolating between these kinetically distinct regimes as a function of the direction normal to the interface.

\begin{figure*}
\includegraphics{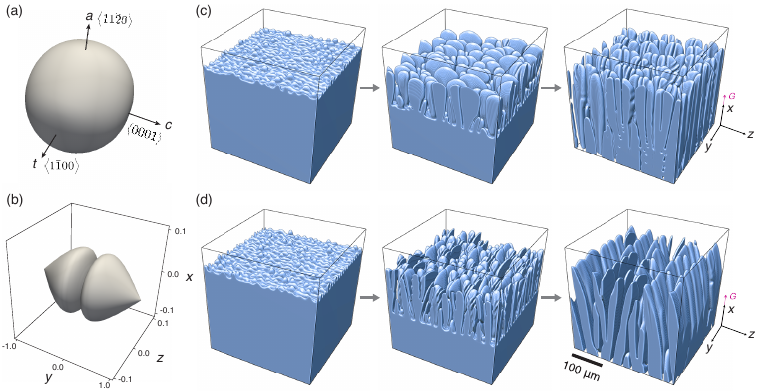}
\caption{\label{fig:figure1}
\added{(a) Numerically calculated equilibrium shape. (b) The angular shape of the kinetic anisotropy in the form $\mu_k^{\left<0001\right>}/\mu_k(\mathbf{n})$. (c)-(d)}
Solid-liquid interfaces captured at $t = 17~\mathrm{s}$ (left), $t = 22~\mathrm{s}$ (center), and $t = 75~\mathrm{s}$ (right) in 3D PF simulations of the directional solidification of a 3 wt.\% aqueous sucrose solution under growth conditions of pulling velocity $V_p = 15~\mathrm{\mu m/s}$ and temperature gradient $G = 12~\mathrm{K/cm}$. Panels (c) and (d) show results with free-energy anisotropy only and with both free-energy and kinetic anisotropies, respectively. In both cases, the $\left<11\bar{2}0\right>$ preferred growth direction is aligned with the temperature gradient $G$, which is parallel to the $x$-axis of the rectangular coordinates, while the $\left<0001\right>$ direction is parallel to the $z$-axis.}
\end{figure*}

Considering a binary water-based system, the liquid-solid phase transformation can be described by a set of sharp-interface equations for standard alloy solidification in the complete-partitioning limit. These include Fickian diffusion of the molecular solute in the liquid, $\partial_{t} c = D \nabla^{2} c$, where $c$ is the solute concentration and $D$ is the diffusivity; the classic Stefan condition for mass conservation at the interface, $c_{l} V_{n} = -\left. D \partial_{n} c \right|^{+}$, where $V_n$ is the normal interface velocity, and $\left. \partial_{n} c \right|^{+}$ is the derivative of the concentration field normal to the interface on the liquid side; and the condition satisfied by the interface temperature $T_I$,
\begin{equation}
T_I = T_M - |m| c_l - \Delta T, \label{sharp_interface}
\end{equation}
where $T_M$ is the melting temperature of the pure substance, $m$ is the slope of the linear liquidus. $\Delta T = \Delta T_c + \Delta T_k$ is the undercooling with contributions from capillarity and interface kinetics given by
\begin{align}
\Delta T_c = \frac{T_{M}}{\Delta h_f} \sum_{i}\left[\Gamma(\mathbf{n})+\frac{\partial^{2} \Gamma(\mathbf{n})}{\partial \theta_{i}^{2}}\right] \frac{1}{R_{i}}, \, \Delta T_k= \frac{V_n}{\mu_k}, \label{Delta_T_c}
\end{align}
respectively, where $\Delta h_f$ is the latent heat of fusion per unit volume, $\Gamma(\mathbf{n})$ is the anisotropic excess interface free energy, $\theta_i$ is the local angle between the normal direction $\mathbf{n}$ and the local principal direction, $R_i$ is the principal radius of curvature, and $\mu_k$ is the atomic attachment kinetic coefficient. The index $i$ in Eq.~\eqref{Delta_T_c} sums over one principal direction in two dimensions (2D) and two perpendicular principal directions in three dimensions (3D). 
To simulate the liquid-solid phase transformation in this binary water-based system, we employ a quantitative PF model that reduces in its thin-interface limit to the sharp-interface equations above, with $\Delta T_k$ in Eq.~\eqref{sharp_interface} being finite on the basal plane normal to the $\left<0001\right>$ directions and negligible on the atomically rough interfaces. Model details, numerical implementation, \added{and material and simulation parameters} are provided in Ref.~\cite{Ji2025QuantitativeTemplating}.

To investigate the separate effects of free-energy and kinetic anisotropies, we first consider the case where the interface is at local thermodynamic equilibrium with $\Delta T_k = 0$ in all directions \cite{Echebarria2004}. For the ice-water interface, $\Gamma(\mathbf{n})$ exhibits six-fold symmetry within the basal plane and two cusps along the $\left<0001\right>$ directions \cite{Ji2025QuantitativeTemplating}. A cusp in $\Gamma(\mathbf{n})$ contributes to the formation of a facet, and its length correlates with the cusp amplitude \cite{Debierre2003Phase-fieldSolidification,Wang2018Phase-fieldGrowth}. However, the free-energy anisotropy of the ice-water interface is weak, and the equilibrium shape only exhibits two small facets normal to the $\left<0001\right>$ directions, as shown in Fig.~\ref{fig:figure1}(a). Here, the coefficients in the $\Gamma(\mathbf{n})$ anisotropy are determined by molecular dynamics models \cite{Davidchack2012IceInteractions,Ji2025QuantitativeTemplating}.
PF simulations with only a weakly anisotropic $\Gamma(\mathbf{n})$ are unable to reproduce the lamellar structures, as seen in Fig.~\ref{fig:figure1}(c). As planar interface breakdown \added{due to perturbations applied to the interface at the beginning of the simulations \cite{Ji2025QuantitativeTemplating}}, Mullins-Sekerka instabilities evolve into columnar cells without prominent facets, growing in a non-steady state with continuous tip splitting. In this scenario, the parity symmetry of the solidification front remains unbroken.

\begin{figure}[htbp!]
\includegraphics{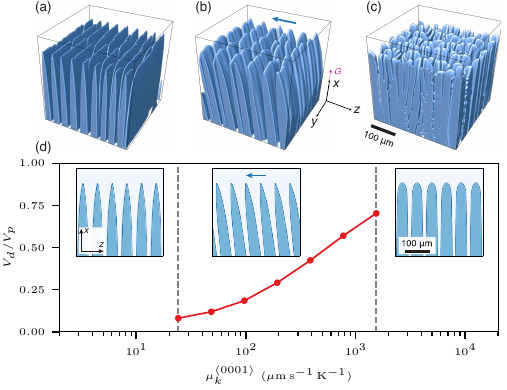}
\caption{\label{fig:figure2_transition} The morphologies of ice crystals in 3D PF simulations with $\mu_k^{\left<0001\right>}=$ 12.1 (a), 41.1 (b), and 775.3 (c) $\mathrm{\mu m/s/K}$. (d) The measured drifting velocity as a function of $\mu_k^{\left<0001\right>}$ from 2D PF simulations. Simulations were performed for the directional solidification of a 3 wt.\% aqueous sucrose solution under growth conditions of $V_p = 15~\mathrm{\mu m/s}$ and $G = 12~\mathrm{K/cm}$. }
\end{figure}


Physically, growth normal to faceted interfaces is significantly slower and is governed by 2D nucleation, layer-by-layer growth, or spiral growth around screw dislocations \cite{Woodruff1973TheInterface}, where $\Delta T_k$ cannot be neglected. The kinetic coefficient $\mu_k^{\left<0001\right>}$ in the $\left<0001\right>$ direction is generally temperature-dependent over a wide range of $\Delta T_k$ \cite{Libbrecht2017}. 
PF simulations can reproduce both linear and nonlinear kinetic relationships for basal plane growth \cite{Ji2025QuantitativeTemplating}, where $\mu_k^{\left<0001\right>}$ is modeled as a finite value and a nonlinear function of $\Delta T_k$ on the basal plane, respectively. For simplicity, in this Letter, we focus on the linear kinetic relationship, which provides a good approximation for the limited range of undercooling within the narrow tip region that governs structural formation.

Incorporating highly anisotropic interface kinetics, as shown in Fig.~\ref{fig:figure1}(b), we perform 3D PF simulations with $\mu_k^{\left<0001\right>} = 41.1 \, \mathrm{\mu m/s/K}$ estimated by fitting experimental measurements \cite{Libbrecht2017}. In these simulations, the $x$-axis is aligned with the temperature gradient and coincides with the $\left[11\bar{2}0\right]$ preferred growth direction, the $z$-axis corresponds to the $\left[0001\right]$ faceted growth direction ($c$-axis), and the $y$-axis corresponds to the $\left[1\bar{1}00\right]$ prism direction ($t$-axis). 
As the planar interface breaks down, morphological instabilities produce small-wavelength protuberances similar to those observed in Fig.~\ref{fig:figure1}(c). However, unlike in Fig.~\ref{fig:figure1}(c), spontaneous parity breaking occurs as the protuberances evolve into columnar cells, forming a partially faceted ice lamellar structure. This behavior qualitatively resembles the morphological instabilities observed during ice-crystal growth in pure undercooled melts \cite{Furukawa1993Three-dimensionalGrowth,Shimada1997PatternWater}.

In Fig.~\ref{fig:figure1}(d), and generally in cases where the temperature gradient axis is contained within the basal plane of ice (the $x$-$y$ plane), the PF evolution equations for solid-liquid interface dynamics are invariant under parity transformation ($z \to -z$), corresponding to mirror reflection about the basal plane. As shown in Fig.~\ref{fig:figure2_transition}, simulations in both 2D and 3D demonstrate spontaneous symmetry breaking within the range $\mu_k^{\mathrm{min}} \leq \mu_k^{\left<0001\right>} \leq \mu_k^{\mathrm{max}}$, resulting in two steady-state growth solutions drifting along the $\pm z$ directions at equal speed while also growing along the $x$ direction. Parity symmetry breaks spontaneously because these solutions, being mirror images of each other, are not invariant under parity transformation.
\added{This range is identified numerically as the interval beyond which the parity-broken solutions disappear at steady state.}
The fact that spontaneous symmetry breaking occurs only within a specific range of the kinetic coefficient can be qualitatively understood by considering the limits of interface kinetics. For very fast kinetics ($\mu_k^{\left<0001\right>} > \mu_k^{\mathrm{max}}$), faceted growth approaches the limit of local equilibrium at the interface. This regime generally does not produce spontaneous symmetry breaking, as observed in standard directional solidification of binary alloys with rough interfaces. In this case, facets are present only in the equilibrium shape but are absent from the growth shape because the interface normal becomes parallel to the $c$-axis only at an infinite distance behind the tip, deep in the mushy zone. 
Conversely, in the limit of very sluggish kinetics ($\mu_k^{\left<0001\right>} < \mu_k^{\mathrm{min}}$), faceted growth becomes too slow, and facets are similarly absent from the growth shape. While the facet drift velocity could theoretically approach zero continuously as $\mu_k^{\left<0001\right>}$ becomes vanishingly small, our results show that tilted growth cannot be sustained below a critical value $\mu_k^{\mathrm{min}}$ of the kinetic coefficient.
Importantly, the experimentally estimated value of $\mu_k^{\left<0001\right>}$ for ice growth falls well within the range $\mu_k^{\mathrm{min}} < \mu_k^{\left<0001\right>} < \mu_k^{\mathrm{max}}$, where spontaneous parity breaking occurs.

\begin{figure}[htbp!]
\includegraphics{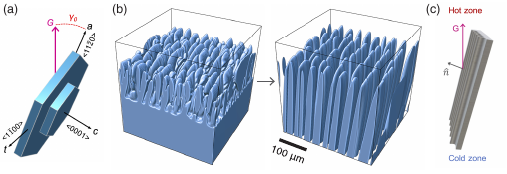}
\caption{\label{fig:figure2}(a) Illustration of the $\gamma_0$ angle between the $a$-axis and the temperature gradient $G$ within the plane containing both the $a$ and $c$ axes. (b) 3D PF simulation of a 3 wt.\% aqueous sucrose solution with $V_p = 15~\mathrm{\mu m/s}$, $G = 12~\mathrm{K/cm}$, and $\gamma_0 = 15^{\circ}$. (c) Unilateral subfeatures on the ice-templated materials tilt towards the hot side of $G$.}
\end{figure}

Parity symmetry can also be broken externally when the basal plane is tilted at an angle $\gamma_0$ relative to the temperature gradient axis, as illustrated in Fig.~\ref{fig:figure2}(a). In this case, the PF evolution equations lose their invariance under the parity transformation ($z \rightarrow -z$). In a 3D PF simulation of a single grain with $\gamma_0 = 15^{\circ}$ [Fig.~\ref{fig:figure2}(b)], columnar cells develop into lamellae with two drifting modes, similar to the case of $\gamma_0 = 0^{\circ}$. However, during competitive growth, ice lamellae associated with one drifting mode dominate, with their facets oriented towards the hot side of the temperature gradient $G$. This results in unilateral substructures on the ice-templated materials tilting towards the hot side of $G$ [Fig.~\ref{fig:figure2}(c)].

\begin{figure}[htbp!]
\includegraphics{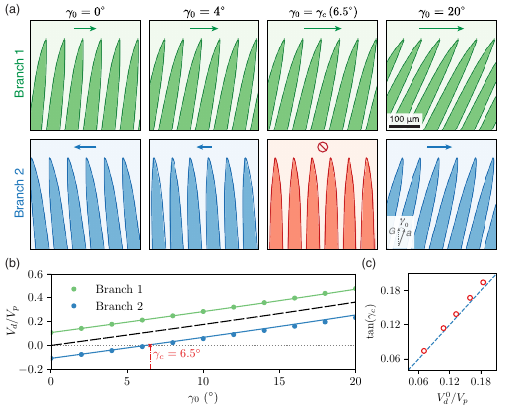}
\caption{\label{fig:figure3} (a) Ice lamellae in two drifting modes at different $\gamma_0$ in 2D PF simulations of a 3 wt.\% aqueous sucrose solution with $V_p = 15~\mathrm{\mu m/s}$ and $G = 12~\mathrm{K/cm}$. The arrows indicate the direction and magnitude of drifting. Drifting ceases at a critical angle $\gamma_c$ in Branch 2. (b) The measured drifting velocity (dots) as a function of $\gamma_0$, where drifting to the right is considered positive.
The black dashed line represents the geometric drifting relation $V_d / V_p = \tan{(\gamma_0)}$, and \added{the solid lines represent the relation in Eq.~\eqref{Vd_pm}.} (c) The measured $\gamma_c$ (dots) agrees with the relation $\tan{(\gamma_c)} = V^0_d / V_p$ (dashed line), where $V^0_d$ is the magnitude of the kinetic drifting at $\gamma_0 = 0^{\circ}$. $\mu_k^{\left<0001\right>}$ is constant ($41.1\,\mathrm{\mu m/s/K}$) in (a)-(b) and varies from $19.4$ to $96.9~\mathrm{\mu m/s/K}$ in (c).}
\end{figure}

To further investigate the orientation selection of primary ice lamellae, we performed PF simulations in 2D, with the $a$ and $c$ axes contained within the simulation domain. \added{As shown in Fig.~\ref{fig:figure3}(a), $\gamma_0$ is progressively increased from two steady-state lamellar solutions at $\gamma_0 = 0^{\circ}$, such that each simulation contains only a single partially faceted structure.} Notably, two distinct branches of lateral drifting modes emerge, with parity symmetry broken in both cases. In Branch 1, ice lamellae drift in the direction of the misorientation (to the right), with their facets oriented in the same direction; the scaled drifting velocity $V_d / V_p$ increases monotonically with $\gamma_0$, as shown in Fig.~\ref{fig:figure3}(b), where drifting to the right is taken as positive. In Branch 2, ice lamellae drift in the opposite direction to the misorientation, with their facets initially oriented in the drifting direction for small $\gamma_0$. The drifting velocity magnitude decreases with $\gamma_0$ until a critical misorientation angle $\gamma_c$ is reached. Remarkably, at $\gamma_0 = \gamma_c$, ice lamellae stop drifting and grow parallel to the temperature gradient axis. This behavior contrasts with the directional solidification of non-faceted materials with atomically rough interfaces, where cellular/dendritic arrays always exhibit lateral drift at nonzero misorientation angles \cite{Akamatsu1997SimilarityCrystals,Deschamps2008,Ghmadh2014,Song2018}. For $\gamma_0 > \gamma_c$ in Branch 2, ice lamellae drift in the direction of the misorientation, and $V_d / V_p$ increases with $\gamma_0$, with facets oriented opposite to the drifting direction.

The drifting of ice lamellae with unbroken parity symmetry in PF simulations for $\mu_k^{\left<0001\right>} < \mu_k^{\mathrm{min}}$ and a finite misorientation angle $\gamma_0$ closely follows the geometric drifting relation ${V_d}/{V_p} = \tan{(\gamma_0)}$ corresponding to $\gamma=\gamma_0$ \cite{SuppMaterial}. This behavior is similar to alloy solidification with a Péclet number $\mathrm{Pe} = \lambda V_p / D \gg 1$ \cite{Akamatsu1997SimilarityCrystals,Deschamps2008,li2012phase,Ghmadh2014,Tourret2015,Song2018}, where $\lambda$ is the primary spacing. However, even though  $\mathrm{Pe} \gg 1$ ($\mathrm{Pe} \approx 6.4$ corresponding to $\lambda=60~\mathrm{\mu m}$) in the simulations in Fig.~\ref{fig:figure3} where parity symmetry is broken, neither Branch 1 nor Branch 2 follows this geometric drifting relation. 
\added{Instead, their drifting can be predicted by the relation
\begin{equation}
V_d = V_p \tan{(\gamma_0) \pm V_d^0}, \label{Vd_pm}
\end{equation}
where the $\pm$ sign corresponds to Branch 1 and 2, respectively, and $V^0_d$ is the drifting velocity magnitude for $\gamma_0 = 0^\circ$ that is entirely controlled by basal plane kinetics \cite{Ji2025QuantitativeTemplating}. As shown in Fig.~\ref{fig:figure3}(b), Eq.~\eqref{Vd_pm} predicts well the $V_d$ for both branches.}
Hence, the drifting of the partially faceted lamellar structure is driven by distinct mechanisms: a \textit{geometric drifting} caused by crystalline misorientation and a \textit{kinetic drifting} induced by basal plane kinetics, \added{where the former is the sole contribution in non-faceted solidification and the latter is unique to faceted systems}.
In Branch 1, the two effects reinforce each other, whereas in Branch 2, they oppose each other. Specifically, in Branch 2, kinetic drifting suppresses geometric drifting for $\gamma_0 < \gamma_c$, leading to drifting in the opposite direction of the misorientation, and the reverse occurs for $\gamma_0 > \gamma_c$.
Since these two effects cancel each other at $\gamma_0 = \gamma_c$, a prediction for $\gamma_c$ can be obtained by simply replacing $\gamma_0$ by $\gamma_c$ in the geometrical relation for unbroken parity symmetry, yielding   
\begin{equation}
\tan{(\gamma_c)} = V^0_d / V_p. \label{gamma_c} 
\end{equation}
As shown in Fig.~\ref{fig:figure3}(c), this prediction agrees well with measurements from PF simulations and further validates our interpretation of two drifting mechanisms.

\begin{figure*}
\includegraphics[width=0.85\textwidth]{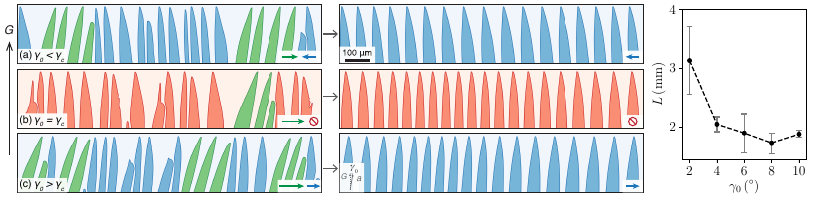}
\caption{\label{fig:figure4} Growth competition between ice lamellae in Branch 1 (green) and Branch 2 (blue) within a single crystal for $\gamma_0=$ $3^{\circ}$ (a), $6.5^{\circ}$ (b), and $10^{\circ}$ (c). Ice lamellae in Branch 2 (red) cease drifting when $\gamma_0 = \gamma_c$. Arrows indicate the directions and magnitudes of drifting for branches of the corresponding color. The middle column shows the dynamically selected Branch 2. The right column presents the solidified length $L$ as a function of $\gamma_0$ when Branch 2 is dynamically selected. Simulations begin from a planar interface at the liquidus temperature during directional solidification of a 3 wt.\% aqueous sucrose solution with $V_p = 15~\mathrm{\mu m/s}$ and $G = 12~\mathrm{K/cm}$.}
\end{figure*}

In spatially extended systems of ice-crystal growth, both branches can coexist within a single grain during the early stages but compete during subsequent growth. This competition is similar to the growth competition of columnar dendritic grains with different crystal misorientations in directional solidification of non-faceted binary alloys \cite{walton1959origin,Tourret2015,Pineau2018,Dorari2021}, albeit involving two different drifting patterns from a single grain instead of two different grains. Here we find that, due to the absence of secondary branches that can influence grain competition in a non-trivial way \cite{Tourret2015,Pineau2018,Dorari2021}, the competition of ice lamellae is well described by the classic Walton and Chalmers minimum undercooling criterion \cite{walton1959origin}, which assumes that the structure with the smallest tip undercooling (i.e., maximum growth temperature) is dynamically selected. For $\gamma_0 = 0^{\circ}$, this criterion predicts that neither of the two structures is preferentially selected since they drift in opposite directions at the same velocity and hence grow with the same undercooling. 
\added{For any finite $\gamma_0$, the criterion predicts that the structure in Branch 2, which drifts at a lower speed [Fig.~\ref{fig:figure3}(b)] and hence has a smaller tip speed $\scriptstyle \sqrt{V_p^2 + V_d^2}$, grows with a smaller undercooling and is therefore dynamically selected, consistent with the results shown in Fig.~\ref{fig:figure4}.}
The right column of Fig.~\ref{fig:figure4} further shows that, for $\gamma_0$ values closer to zero, the selection process requires a longer solidified length $L$ since the difference in undercooling between the two structures is small \added{(i.e., Branch 2 has less advantage)} and the process exhibits greater stochasticity, as indicated by the larger standard error of the mean according to five simulations per data point. For $\gamma_0$ near or larger than $\gamma_c$, the competition process typically finishes over a similar distance of approximately $L \approx 1.8~\mathrm{mm}$.
This selection mechanism leads to unilateral features on ice-templated materials tilting towards the hot side of $G$, a characteristic of Branch 2 for $\gamma_0 > \gamma_c$. Given that $\gamma_c$ is generally small (a few degrees) according to Eq.~\eqref{gamma_c}, Branch 2 lamellae with $\gamma_0 > \gamma_c$ are expected to dominate in spatially extended experiments \cite{kamm2023x} with several grains of varying crystalline misorientation.

In conclusion, our phase-field simulation results demonstrate that the tilted growth of ice lamellar arrays during ice templating can be understood within the framework of spontaneous parity breaking. \added{Firstly, the Mullins-Sekerka instabilities and spontaneous parity breaking give rise to asymmetric structures in two drifting modes that initially coexist and subsequently grow competitively.} Secondly, simulations show that parity breaking only occurs over a range of the kinetic coefficient $\mu_k^{\mathrm{min}} \leq \mu_k^{\left<0001\right>} \leq \mu_k^{\mathrm{max}}$ that controls the growth of ice facets along the $c$-axis. Importantly, this range is wide and encompasses the value of $\mu_k^{\left<0001\right>}$ estimated from experimental measurements. Thirdly, they reveal that, when a small misorientation angle $\gamma_0$ between the $a$-axis of the ice crystal and the temperature gradient axis is present, the two branches of parity-broken structures become non-equivalent. They drift at different velocities, in opposite direction when $\gamma_0<\gamma_c$ or the same direction when $\gamma_0>\gamma_c$, where $\gamma_c$ (typically a few degrees) can be predicted from a simple geometrical relation. Finally, the results show that the tilted lamellar structure with the smallest drifting velocity (the Branch 2 structure) is dynamically selected consistent with the minimum undercooling criterion. The latter result provides a natural explanation for why templated structures, which can be generally expected to be formed by ice crystals with $\gamma_0>\gamma_c$ (i.e., $\gamma_0$  larger than a few degrees) in large polycrystalline samples, are commonly observed to be oriented toward the hot side.

This work was supported through NASA grants 80NSSC18K0305 and 80NSSC21K0039. Numerical simulations were performed on Northeastern University's Discovery cluster located in Massachusetts Green High Performance Computing Center (MGHPCC) in Holyoke, MA. K.J. acknowledges that this work was partially supported by Lawrence Livermore National Laboratory (LLNL) under Contract DE-AC52-07NA27344. The authors thank Ulrike Wegst for valuable discussions linked to experimental observations.


\bibliography{references}

\end{document}